\documentclass[12pt]{article}
\usepackage{graphics}
\usepackage{epsf}
\begin{document}
\def\stop{\tilde{t}}
\def\sto{\tilde{t}_1}
\def\stt{\tilde{t}_2}
\def\stl{\tilde{t}_L}
\def\str{\tilde{t}_R}
\def\msto{m_{\sto}}
\def\mstosq{m_{\sto}^2}
\def\mstt{m_{\stt}}
\def\msttsq{m_{\stt}^2}
\def\mt{m_t}
\def\mtsq{m_t^2}
\def\sint{\sin\theta_{\stop}}
\def\cost{\cos\theta_{\stop}}
\def\sintsq{\sin^2\theta_{\stop}}
\def\costsq{\cos^2\theta_{\stop}}
\def\sinb{\sin\beta}
\def\cosb{\cos\beta}
\def\sinbb{\sin (2\beta)}
\def\cosbb{\cos (2 \beta)}
\def\tgb{\tan \beta}
\def\mqtt{\M_{\tilde{Q}_3}^2}
\def\mutt{\M_{\tilde{U}_{3R}}^2}
\def\sbottom{\tilde{b}}
\def\sbo{\tilde{b}_1}
\def\sbt{\tilde{b}_2}
\def\sbl{\tilde{b}_L}
\def\sbr{\tilde{b}_R}
\def\msbo{m_{\sbo}}
\def\msbosq{m_{\sbo}^2}
\def\msbt{m_{\sbt}}
\def\msbtsq{m_{\sbt}^2}
\def\mt{m_t}
\def\mtsq{m_t^2}
\def\selectron{\tilde{e}}
\def\seo{\tilde{e}_1}
\def\set{\tilde{e}_2}
\def\sel{\tilde{e}_L}
\def\ser{\tilde{e}_R}
\def\mseo{m_{\seo}}
\def\mseosq{m_{\seo}^2}
\def\mset{m_{\set}}
\def\msetsq{m_{\set}^2}
\def\me{m_e}
\def\mesq{m_e^2}
\def\snue{\tilde{\nu_e}}
\def\set{\tilde{e}_2}
\def\snul{\tilde{\nu}_L}
\def\msnue{m_{\snue}}
\def\msnuesq{m_{\snue}^2}
\def\smuon{\tilde{\mu}}
\def\smul{\tilde{\mu}_L}
\def\smur{\tilde{\mu}_R}
\def\msmul{m_{\smul}}
\def\msmulsq{m_{\smul}^2}
\def\msmur{m_{\smur}}
\def\msmursq{m_{\smur}^2}
\def\stau{\tilde{\tau}}
\def\stauo{\tilde{\tau}_1}
\def\staut{\tilde{\tau}_2}
\def\staul{\tilde{\tau}_L}
\def\staur{\tilde{\tau}_R}
\def\mstauo{m_{\stauo}}
\def\mstauosq{m_{\stauo}^2}
\def\mstaut{m_{\staut}}
\def\mstautsq{m_{\staut}^2}
\def\mtau{m_\tau}
\def\mtausq{m_\tau^2}
\def\gluino{\tilde{g}}
\def\mgluino{m_\tilde{g}}
\def\neuto{\tilde{\chi}_1^0}
\def\neutt{\tilde{\chi}_2^0}
\def\neutth{\tilde{\chi}_3^0}
\def\neutf{\tilde{\chi}_4^0}
\def\charop{\tilde{\chi}_1^+}
\def\chargtp{\tilde{\chi}_2^+}
\def\charom{\tilde{\chi}_1^-}
\def\chargtm{\tilde{\chi}_2^-}
\def\bino{\tilde{b}}
\def\wino{\tilde{w}}
\def\photino{\tilde{\gamma}}
\def\zino{tilde{z}}
\def\sdowno{\tilde{d}_1}
\def\sdownt{\tilde{d}_2}
\def\sdownl{\tilde{d}_L}
\def\sdownr{\tilde{d}_R}
\def\supo{\tilde{u}_1}
\def\supt{\tilde{u}_2}
\def\supl{\tilde{u}_L}
\def\supr{\tilde{u}_R}

\newcommand{\be}{\begin{equation}}
\newcommand{\beq}{\begin{equation}}
\newcommand{\eeq}{\end{equation}}
\newcommand{\ee}{\end{equation}}

\newcommand{\beqn}{\begin{eqnarray}}
\newcommand{\eeqn}{\end{eqnarray}}
\newcommand{\bea}{\begin{eqnarray}}
\newcommand{\ena}{\end{eqnarray}}
\newcommand{\ra}{\rightarrow}
\newcommand{\susy}{{\cal SUSY$\;$}}
\newcommand{\su}{$ SU(2) \times U(1)\,$}

\newcommand{\gag}{$\gamma \gamma$ }
\newcommand{\gam}{\gamma \gamma }
\def\W{{\mbox{\boldmath $W$}}}
\def\B{{\mbox{\boldmath $B$}}}
\newcommand{\np}{Nucl.\,Phys.\,}
\newcommand{\pl}{Phys.\,Lett.\,}
\newcommand{\pr}{Phys.\,Rev.\,}
\newcommand{\prl}{Phys.\,Rev.\,Lett.\,}
\newcommand{\prep}{Phys.\,Rep.\,}
\newcommand{\zp}{Z.\,Phys.\,}
\newcommand{\sovjnp}{{\em Sov.\ J.\ Nucl.\ Phys.\ }}
\newcommand{\nuclinst}{{\em Nucl.\ Instrum.\ Meth.\ }}
\newcommand{\annp}{{\em Ann.\ Phys.\ }}
\newcommand{\intjmp}{{\em Int.\ J.\ of Mod.\  Phys.\ }}

\newcommand{\eps}{\epsilon}
\newcommand{\mw}{M_{W}}
\newcommand{\mww}{M_{W}^{2}}
\newcommand{\mwmw}{M_{W}^{2}}
\newcommand{\mhmh}{M_{H}^2}
\newcommand{\mz}{M_{Z}}
\newcommand{\mzz}{M_{Z}^{2}}

\newcommand{\lra}{\leftrightarrow}
\newcommand{\tr}{{\rm Tr}}
\def\ls1{{\not l}_1}
\newcommand{\cms}{centre-of-mass\hspace*{.1cm}}

\newcommand{\dkg}{\Delta \kappa_{\gamma}}
\newcommand{\dkz}{\Delta \kappa_{Z}}
\newcommand{\dz}{\delta_{Z}}
\newcommand{\dgz}{\Delta g^{1}_{Z}}
\newcommand{\dgzt}{$\Delta g^{1}_{Z}\;$}
\newcommand{\la}{\lambda}
\newcommand{\lag}{\lambda_{\gamma}}
\newcommand{\lambdae}{\lambda_{e}}
\newcommand{\laz}{\lambda_{Z}}
\newcommand{\lnl}{L_{9L}}
\newcommand{\lnr}{L_{9R}}
\newcommand{\lt}{L_{10}}
\newcommand{\lu}{L_{1}}
\newcommand{\ld}{L_{2}}
\newcommand{\cw}{\cos\theta_W}
\newcommand{\sw}{\sin\theta_W}
\newcommand{\tw}{\tan\theta_W}
\def\cww{\cos^2\theta_W}
\def\sww{\sin^2\theta_W}
\def\tww{\tan^2\theta_W}

\newcommand{\epm}{$e^{+} e^{-}\;$}
\newcommand{\epemt}{$e^{+} e^{-}\;$}
\newcommand{\epem}{e^{+} e^{-}\;}
\newcommand{\ememt}{$e^{-} e^{-}\;$}
\newcommand{\emem}{e^{-} e^{-}\;}
\newcommand{\eeww}{e^{+} e^{-} \ra W^+ W^- \;}
\newcommand{\eewwt}{$e^{+} e^{-} \ra W^+ W^- \;$}
\newcommand{\epemww}{e^{+} e^{-} \ra W^+ W^- }
\newcommand{\epemwwt}{$e^{+} e^{-} \ra W^+ W^- \;$}
\newcommand{\eennhht}{$e^{+} e^{-} \ra \nu_e \bar \nu_e HH\;$}
\newcommand{\eennhh}{e^{+} e^{-} \ra \nu_e \bar \nu_e HH\;}
\newcommand{\ppwg}{p p \ra W \gamma}
\newcommand{\wwhh}{W^+ W^- \ra HH\;}
\newcommand{\wwhht}{$W^+ W^- \ra HH\;$}
\newcommand{\ppwz}{pp \ra W Z}
\newcommand{\ppwgt}{$p p \ra W \gamma \;$}
\newcommand{\ppwzt}{$pp \ra W Z \;$}
\newcommand{\gamgamt}{$\gamma \gamma \;$}
\newcommand{\gamgam}{\gamma \gamma \;}
\newcommand{\egamt}{$e \gamma \;$}
\newcommand{\egam}{e \gamma \;}
\newcommand{\gamgamwwt}{$\gamma \gamma \ra W^+ W^- \;$}
\newcommand{\gamgamwwht}{$\gamma \gamma \ra W^+ W^- H \;$}
\newcommand{\gamgamwwh}{\gamma \gamma \ra W^+ W^- H \;}
\newcommand{\gamgamwwhht}{$\gamma \gamma \ra W^+ W^- H H\;$}
\newcommand{\gamgamwwhh}{\gamma \gamma \ra W^+ W^- H H\;}
\newcommand{\ggww}{\gamma \gamma \ra W^+ W^-}
\newcommand{\ggwwt}{$\gamma \gamma \ra W^+ W^- \;$}
\newcommand{\ggwwht}{$\gamma \gamma \ra W^+ W^- H \;$}
\newcommand{\ggwwh}{\gamma \gamma \ra W^+ W^- H \;}
\newcommand{\ggwwhht}{$\gamma \gamma \ra W^+ W^- H H\;$}
\newcommand{\ggwwhh}{\gamma \gamma \ra W^+ W^- H H\;}
\newcommand{\ggwwz}{\gamma \gamma \ra W^+ W^- Z\;}
\newcommand{\ggwwzt}{$\gamma \gamma \ra W^+ W^- Z\;$}
\def\smx{{\cal{S}} {\cal{M}}\;}

\newcommand{\ptu}{p_{1\bot}}
\newcommand{\vecptu}{\vec{p}_{1\bot}}
\newcommand{\ptd}{p_{2\bot}}
\newcommand{\vecptd}{\vec{p}_{2\bot}}
\newcommand{\ie}{{\em i.e.}}
\newcommand{\cm}{{{\cal M}}}
\newcommand{\cl}{{{\cal L}}}
\newcommand{\cd}{{{\cal D}}}
\newcommand{\cv}{{{\cal V}}}
\def\slashc{c\kern -.400em {/}}
\def\slashL{L\kern -.450em {/}}
\def\slashcl{\cl\kern -.600em {/}}
\def\Ww{{\mbox{\boldmath $W$}}}
\def\B{{\mbox{\boldmath $B$}}}
\def\noi{\noindent}
\def\nn{\noindent}
\def\sm{${\cal{S}} {\cal{M}}\;$}
\def\nph{${\cal{N}} {\cal{P}}\;$}
\def\sb{$ {\cal{S}}  {\cal{B}}\;$}
\def\ssb{${\cal{S}} {\cal{S}}  {\cal{B}}\;$}
\def\ssbe{{\cal{S}} {\cal{S}}  {\cal{B}}}
\def\cviol{${\cal{C}}\;$}
\def\pviol{${\cal{P}}\;$}
\def\cpviol{${\cal{C}} {\cal{P}}\;$}

\newcommand{\lgg}{\lambda_1\lambda_2}
\newcommand{\lww}{\lambda_3\lambda_4}
\newcommand{\ppin}{ P^+_{12}}
\newcommand{\pmin}{ P^-_{12}}
\newcommand{\ppout}{ P^+_{34}}
\newcommand{\pmout}{ P^-_{34}}
\newcommand{\sinsq}{\sin^2\theta}
\newcommand{\cossq}{\cos^2\theta}
\newcommand{\yt}{y_\theta}
\newcommand{\hppll}{++;00}
\newcommand{\hpmll}{+-;00}
\newcommand{\hpplt}{++;\lambda_30}
\newcommand{\hpmlt}{+-;\lambda_30}
\newcommand{\hpptt}{++;\lambda_3\lambda_4}
\newcommand{\hpmtt}{+-;\lambda_3\lambda_4}
\newcommand{\dk}{\Delta\kappa}
\newcommand{\klam}{\Delta\kappa \lambda_\gamma }
\newcommand{\kac}{\Delta\kappa^2 }
\newcommand{\lac}{\lambda_\gamma^2 }
\def\gamgamtzz{$\gamma \gamma \ra ZZ \;$}
\def\gamgamtww{$\gamma \gamma \ra W^+ W^-\;$}
\def\gamgamtwwe{\gamma \gamma \ra W^+ W^-}

\def\mh{M_h}
\def\sintt{\sin 2\theta_{\stop}}
\def\cosbsq{\cos^2\theta_{\sbo}}
\def\sinbsq{\sin^2\theta_{\sbo}}
\def\mbsq{m_b^2}
\bibliographystyle{unsrt}
\begin{titlepage}
\def\baselinestretch{1.2}
\topmargin     -0.25in

\vspace*{\fill}
\begin{center}
{\large {\bf Z radiation off stops at a linear collider }} 
\vspace*{0.5cm} 

\begin{tabular}[t]{c}

{\bf G.~B\'elanger$^{1}$, F.~Boudjema$^{1}$,  T.Kon$^{2}$ and 
V~.~Lafage$^{3}$ } 
 \\
\\
{\it 1. Laboratoire de Physique Th\'eorique}
{\large LAPTH}
\footnote{URA 14-36 du CNRS, associ\'ee  \`a
l'Universit\'e de Savoie.}\\
 {\it Chemin de Bellevue, B.P. 110, F-74941 Annecy-le-Vieux,
Cedex, France.}\\

{\it 2. Seikei University, Musashino, Tokyo 180-8633, Japan. } \\ 

{\it 3. High Energy Accelerator Research Organisation, KEK,}\\ 
{\it Tsukuba, Ibakari 305-801, Japan.}\\

\end{tabular}
\end{center}

\centerline{ {\bf Abstract} }
\baselineskip=14pt
\noindent
 {\small We calculate  $\epem\to \sto \sto Z$ at a linear collider.
 For large splitting between the two stops the cross-section is sensitive to the
 value of $m_{\stt}$  when this particle is too heavy to be directly 
 produced. The results are compared to $\epem\to \sto \sto h$  }
\vspace*{\fill}


\vspace*{0.1cm} 
\rightline{LAPTH-740/99} 
\rightline{KEK-CP-086/99}
\rightline{AP-SU-99/02}
\rightline{{\large  June 1999}} 
\end{titlepage}
\baselineskip=18pt

\setcounter{section}{0} \setcounter{subsection}{0} 
\setcounter{equation}{0} 
\def\thesubsection {\thesection.\arabic{subsection}}
\def\theequation{\thesection.\arabic{equation}}

\section{Introduction}

 In the minimal 
supersymmetric standard model (MSSM),the third generation of 
sfermions   plays a special role both from the 
theoretical and phenomenological point of view. 
 Large mixing in the third generation 
can induce large splitting between left and right-handed squarks 
leading in particular  to a top squark significantly lighter than 
other sfermions. With the Higgs, the stop could be the lightest 
scalar of the MSSM and thus particularly interesting to study at a 
linear collider where the moderate total energy restricts  the 
number of sparticles that can be directly produced.

 A large mixing in the stop sector not only 
drives the lightest stop mass down but also can induce large 
couplings between the top squark and the Higgs affecting in many 
ways the phenomenology of the Higgs. First,  radiative corrections 
due to top and stop  can significantly shift the value of the 
tree-level  mass of  the 
Higgs\cite{RChiggsmass_oneloop,higgsmass_twoloop_exact}. More 
importantly, the Higgs signals at LHC-Tevatron could be completely  
different from what is generally expected in the MSSM with no 
mixing. The main discovery channel at the LHC, the 
loop induced direct production $gg\to h\to \gamma\gamma$, can be 
severely suppressed\cite{global_hgg_lhc}.  Furthermore one 
expects modification of  the two-photon width of the Higgs
and possibly a large
 cross-section for associated Higgs production $\sto \sto h$ or $\stt \sto h$
 \cite{nous_Rggstophiggs_lhc,stophiggs_LHC,moretti_stophiggs,abdel_t1t1hlc},  
 where $\sto(\stt)$ is the lightest(heaviest) top squark.

From the theoretical point of view there is also ample motivation 
for considering scenarios of light third generation sfermions. For 
example   in inverted hierarchy models only sfermions of the third 
generation are light enough to be accessible at LHC/Tevatron or a 
future linear collider, all  others being above the TeV
scale\cite{Inverted_hierarchy,Inverted_hierarchy2}. 
Even in models where one assumes universality of sfermion masses 
at a high scale, the degeneracy is lifted once the masses are run 
down to the weak scale according to the renormalization group 
equations and a light $\sto$ is obtained particularly in models with non-negligible trilinear 
couplings. These models are especially attractive since they solve the supersymmetric flavor problem while preserving the naturality argument. Another motivation
 for considering a light stop is the  possibility of obtaining 
 electroweak baryogenesis\cite{Stop_Baryogenesis}.

  In scenarios with a light stop, as was pointed out in \cite{Desy_PhysRep,
  stop_Sopczak,stop_Sopczak2},
   the stop pair 
production at   a polarized linear collider can provide a 
measurement of both the stop mass   and the mixing angle. 
Provided sufficient phase space it was also pointed out that the associated 
production of stops with a Higgs ($\epem\to \sto\sto h$) could be
 observable at a high energy linear collider for 
some region of the parameter space \cite{abdel_t1t1hlc,eenous_t1t1h}. In fact in the presence of 
mixing (associated with a large trilinear term $A_t$) and a heavy $\stt$, the 
coupling of the Higgs to $\sto$ becomes very large. In \cite{
eenous_t1t1h} we advocated using the information from $\sto \sto 
h$ combined with the measurement of $\mh$ to extract the value of  
$\tan\beta$ and $m_{\stt}$ while the $\stt$ would be too heavy 
to be directly produced through $\epem \to \sto\stt$. This is possible 
since to a good approximation we have shown that apart from 
$\tan\beta$ the $\sto \sto h$ vertex depends 
 only the parameters of the stop sector and so do the dominant corrections
 to $\mh$\cite{eenous_t1t1h}. 
  However when $\sto\sto h$ is kinematically 
accessible so is $\sto\sto Z$. The latter process also contains a 
diagram with Higgs exchange and is therefore also sensitive to the 
value of  the $\sto \sto h$ coupling. The purpose of this paper is to show that
 although  the dependence on the $\sto \sto h$ coupling is 
milder than in $\sto \sto h$ production,  the $\sto\sto Z$ process 
features in general a larger cross-section and  it can 
provide complementary information on the parameters of the stop 
sector.

\section{Stop parameters}

The stop sector involves three independent parameters that can be 
taken as the physical masses of the two squarks and the mixing 
angle. The stop mass eigenstates are defined through the mixing 
angle $\theta_{\tilde{t}}$, with  the lightest stop, $\sto$, 
\beq 
\sto=\cost \; \stl + \sint \; \str
 \eeq
 The mixing angle is 
related to the off-diagonal term of the mass matrix  

\beqn \label{s2t} \sin (2 \theta_{\stop})= \frac{2 \; 
m_{\tilde{t}_{LR}}^2}{\mstosq-\msttsq} 
=\frac{ -2\mt(A_t+\mu/\tgb)}{\mstosq-\msttsq}\eeqn 

\noi with  $A_t$ 
the trilinear parameter of the top and $\mu$ the Higgs mixing parameter.  

When only one stop is kinematically accessible as would most 
likely be the case at the linear collider,  stop pair production ($\sto\sto$)
 allows for the determination of one mass, $\msto$. The 
cross-section featuring a strong dependence on $\costsq$, the 
amount of mixing can also  be determined. This can best be done 
using polarized beams.  A precision  
 at  the  percent level has  been estimated  for the
high-luminosity 500GeV
collider.\cite{viennamixing}

 In the decoupling limit of 
large $M_A$,{\footnote{See\cite{eenous_t1t1h} for further 
discussion on the validity of this approximation} 
 it has 
been shown \cite{eenous_t1t1h} that the $\sto \sto h$ vertex 
depends only on the three parameters of the stop sector together 
with $\tan\beta$, 

\beqn \label{approxtth} 
 V_{\sto \sto h} &\simeq &\frac{g}{\mw} \biggl( \sin^2(2 \theta_{\stop})
 \frac{(\mstosq-\msttsq)}{4} \;+\; \mt^2 \nonumber \\
 &+& \mzz \cos (2
\beta)\left((\frac{1}{2}-\frac{2}{3} \sww)\costsq +\frac{2}{3} 
\sww \sintsq \right) \biggr) \eeqn 

\noi Note that in this approximation, there is no  $\mu$ dependence in the vertex and  
that the $\tgb$ dependence  arises from the small D-term. 
For both $\sto\sto Z$ and $\sto\sto h$  processes,  the value of  $\tgb$ affects mainly the 
computation of the Higgs mass.

 The vertex almost vanishes when the stop/top contributions cancel 
each other. This occurs at 
\beqn \sin(2 \theta_{\stop})\approx 
\frac{4\mt^2}{m_{\stt}^2-m_{\sto}^2} \eeqn

\noi  At small values of $\sin 2\theta_{\stop}$, 
 the $\sto \sto h$ vertex, up to small D-terms is of the same order as
 the $t\overline{t}h$ vertex since it is dominated by the $\mt^2$ term 
 in \ref{approxtth}.
For large values of the $\sto \sto h$ vertex, the cross section
for $\epem \to \sto\sto Z$ 
gets quite large. This occurs for maximal mixing, $\sin 2\theta_{\stop}\approx 1$, with a large 
splitting between the two stop physical masses, $\mstt \gg \msto$. However it is 
precisely for this configuration that one has some strong 
constraints. These will be discussed in the next section.

\section{Constraints from $\mh$, $\Delta\rho$ and CCB}

The most stringent constraint generally arises from $\Delta\rho$ which 
receives contributions from both sbottom and stops. When there is 
a large splitting between the masses of squarks, the contribution 
to the gauge-boson self energies becomes sizable and grows with the 
mass of the heavier squark.  The soft-breaking mass,  
$m_{\tilde{Q}_L}$, being common to  the two members of the SU(2) 
doublet, one parameter of the sbottom sector is related to that
of the stop sector: 
\beqn\label{msu2}  m^2_{\tilde{Q_L}}&=&\costsq \mstosq+\sintsq 
\msttsq -\mtsq - M_Z^2\cosbb (\frac{1}{2}-\frac{2}{3}s^2_W)\\ & 
=&\cosbsq \msbosq+\sinbsq m_{\tilde{b}_2}^2 -\mbsq - M_Z^2\cosbb 
(-\frac{1}{2}+\frac{1}{3}s^2_W) \eeqn 

\noi If we restrict ourselves to the limit of small 
mixing in the sbottom sector, $\theta_b=0$,  we are  left with three free parameters among the 
 five parameters of the third generation squark sector. These 
will be taken as the physical masses of the stops and the mixing 
angle, $\theta_{\stop}$ . 
In this limit $\sbo\approx \sbl$ and is the 
only component entering the radiative corrections to $\Delta\rho$. 
The $\sbt$ is now purely $\sbr$ and  decouples from the 
constraints. 
There are essentially three contributions to $\Delta\rho$, which in the limit 
of small mixing in the sbottom sector simplifies to,

\beqn\label{deltarho} \Delta\rho=-\sintsq\costsq f(\msto,\mstt) 
+\costsq f(\msto,m_{\tilde{Q}_L})+\sintsq f(\mstt,m_{\tilde{Q}_L}) 
\eeqn 

\noi where the functions $f(m_1,m_2)$ include both one- and 
two-loop corrections and are defined in \cite{Drhosusy_2loop}. They 
vanish for equal masses.

\begin{figure*}[htbp]
\begin{center}
\mbox{\epsfxsize=14cm\epsfysize=12cm\epsffile{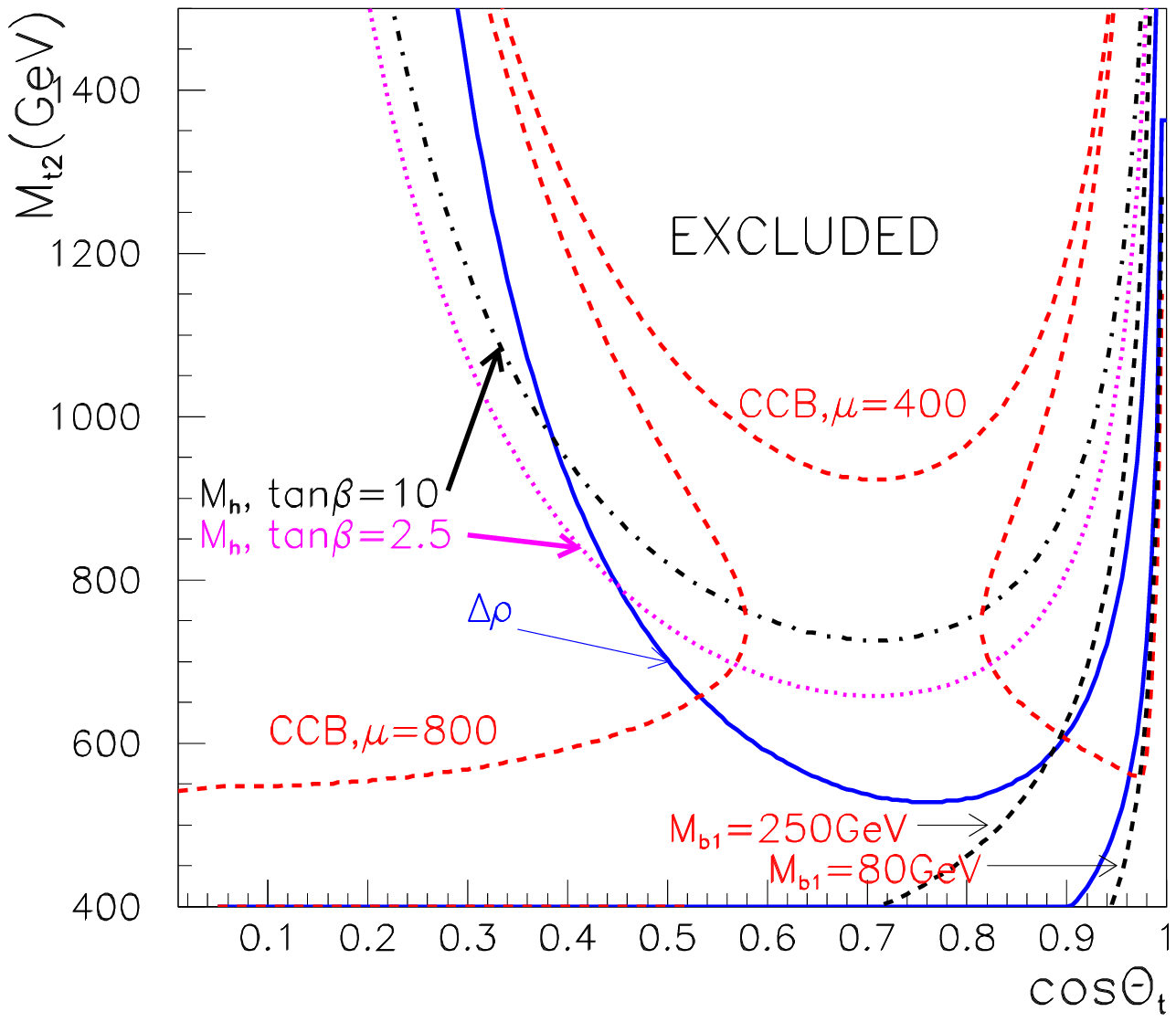}} 
\vspace*{-1cm}
\caption{\label{constraint}{\em  Constraint from $\Delta\rho\le 0.0013$(full 
line), $\mh\ge 90GeV$(dash-dot), CCB(dash) and $\msbo$  for $\tgb=10, 
\mu=400 GeV$, $m_{\sto}=120GeV$ and 
$M_A=1$TeV. The $\mh$ constraint for $\tgb=2.5$ is also shown 
(dot). The excluded region determined by the above constraints is within the respective boundaries indicated. Note that for  $\cost\approx 1$, 
the $\Delta\rho$ constraint also excludes the region to the right of the second branch of the
 $\Delta\rho$ curve where the present limit on the mass of the sbottom is contained.   Requiring sbottom production to be above threshold at a 500GeV linear collider ($\msbo\ge 250GeV$) excludes the region to the right of the curve.The CCB 
constraint for $\mu=800 GeV$ is also displayed, the excluded region 
lies between the two  CCB, $\mu=800$, curves.}} 
\end{center}
\end{figure*}

Imposing the constraint that $\Delta\rho\le 0.0013$\cite{Langacker_Fits98}, we found, as shown in
Fig.\ref{constraint}, that for  large mixing $\sintt\approx 1$, 
the large values of $m_{\stt}$ are ruled out. These results assume a fixed value of $\msto=120GeV$. 
For a near maximal mixing angle, the $\stt$ cannot 
exceed 542GeV while for a mixing $\cos\theta_{\stop}\approx .4$ one can 
allow $\stt$ up to 900GeV.
When $\cos\theta_{\stop}$ is 
small, ($\sin 2\theta_{\stop}\approx 0$) masses in excess of 1TeV are allowed as  the contributions from the 
terms with large mass splittings are damped by the factor 
$\sintsq$.
 When $\cos\theta_{\stop}\approx 1$ there exist both a 
lower and upper limit on $m_{\stt}$. The region where  $m_{\stt}$ 
is small corresponds to one where the common SU(2) squark mass is 
very low, {\footnote{Note that when $\cos\theta_{\stop}\approx 1$,
 the sbottom mass drops below the direct experimental lower 
bound.} } all terms give a significant contribution to 
$\Delta\rho$. It is only when $m_{\stt}$ increases that 
$m_{\tilde{Q_L}}\approx m_{\sto}$, due to the near  degeneracy in 
mass,  this term does not contribute to $\Delta\rho$. Furthermore  
there is a near cancellation between the two contributions 
involving the $\stt$.

A large $\sto\sto h$ vertex also means an important contribution to the Higgs 
mass. We have taken the approximate formulae at 
one-loop\cite{RChiggsmass_Ellis} including a running top mass to 
incorporate the leading two-loop corrections. In fact the 
correction to the Higgs mass depends on exactly the same 
combination of parameters than the one entering the $\sto \sto h$ 
vertex\cite{eenous_t1t1h}. For large mixings and large $\stt$ mass, the 
Higgs mass is driven below the present direct experimental limit, $\mh\le
90GeV$, 
and as the $\stt$ mass increases is rapidly driven negative. While the 
value of the Higgs mass is dependent on $\tgb$, there is only a 
small shift in the allowed region as $\mh$ drops very rapidly when 
the mixing increases.  For the region of large $\sintt$, the 
constraint from $\Delta\rho$ is always more stringent, it is only 
for mixings below $\approx .4$ that the Higgs mass becomes the 
most stringent constraint. 

One should also mention the constraint arising
 from the 
requirement that the parameters  do not induce colour and charge 
breaking global minima (CCB)\cite{CCBnaive}. An upper bound on 
$A_t$, or on the amount of mixing,  follows from this requirement. 
However it has been argued that the constraints based on the 
global minima may be too restrictive\cite{CCB_Kusenko}. It was 
shown that for a wide range of parameters, the global CCB minimum 
becomes irrelevant on the ground that the time required to reach 
the lowest energy state exceeds the present age of the universe. 
Taking the tunneling rate into account results in a milder 
constraint which may be approximated\cite{CCB_Kusenko} by : 
\beqn \label{CCBkusenko} A_t^2 +3\mu^2 < 7.5 (m_{\tilde{Q}_L}^2 + 
m_{\stop_R}^2) \eeqn

\noi 
This constraint depends on $\mu$ both explicitly and in the 
calculation of $A_t$ in terms of physical parameters(see \ref{s2t}). 
 For the parameters we are entertaining here, with an intermediate value for $\mu$, the mild CCB 
constraint does not come into effect, it is always superseded by both 
the $\Delta\rho$ and $\mh$ constraints. This value of $\mu$ was chosen such that there would not be
 other supersymmetric particles such as gauginos directly produced at the LC. 
However for large values of  
 $|\mu|$ this constraint can become very relevant as both an
 upper limit and a lower limit on $m_{\stt}$ are obtained.  In fact
 for $\mu=800GeV$ the whole area of near maximal mixing is 
ruled out for any values of $m_{\stt}$.  Note that in the region near 
$\cost=1$ the lower bound on $m_{\stt}$ increases significantly, 
in this region one obtains negative $m^2_{\tilde{Q}_L}$ inducing 
CCB.  Both the 
curves for $\mu=400GeV$ and $800GeV$ are displayed in 
Fig.~\ref{constraint}.

 Although the sbottom  mass does not enter the calculation of the $\sto \sto Z$, 
 one has to make sure that the sbottom mass does not drop 
 below the experimental direct bound of roughly 80GeV. 
 This can occur in the region where $\cost\approx 1$ especially for
 the low values of $m_{\stt}$. However this constraint is also 
 superseded by the $\Delta\rho$ constraint, Fig.\ref{constraint}. 
Although not strictly a constraint,  we are also interested in 
knowing whether or not the 
 sbottom is light enough to be directly pair-produced at the
  linear collider. If 
 such is the case, the direct measurement of its mass, at least in the 
 approximation of small mixing, would be sufficient to completely define
 the parameters of the stop sector. 
 Note that the region where $\sbo$ is light enough to be 
 pair-produced corresponds to either small $m_{\stt}$ or 
 $\cost\approx 1$. In either case the $\sto\sto h$ vertex is not 
 large as seen in Fig.\ref{contth}. 
 
 \begin{figure*}[htbp]
\begin{center}
\mbox{\epsfxsize=14cm\epsfysize=12cm\epsffile{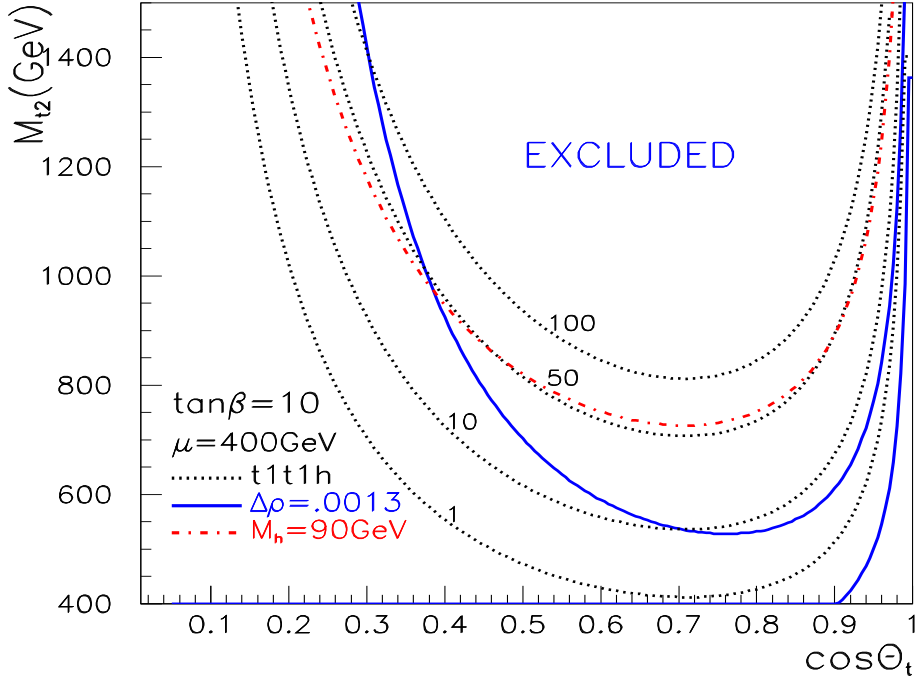}} 
\caption{\label{contth}{\em  Equipotential lines (dotted) for the normalized 
coupling $R_{\sto}=1,10,50,100$, see text,  with  $\tgb=10$ and $\mu=400 
GeV$. The exclusion regions corresponding to $\Delta\rho\le .0013$ and $\mh\le 
90GeV$ are reproduced from Fig.\ref{constraint}.}} 
\end{center}
\end{figure*}
 
 As we are interested in probing the large Yukawa coupling, it is 
 useful to estimate the strength of the $\sto \sto h$ 
 coupling before going to the full calculation. To this end we 
 define the   coupling squared normalized to the coupling in the no-mixing limit
 and without a D-term, this corresponds approximately to the
 strength of the $t\overline{t}h$ coupling,
 \beqn  
  R_{\sto }= \left(\frac{\mw V_{\sto\sto 
 h}}{ g \mt^2 }\right)^2
 \eeqn
 In Fig.~\ref{contth} we show contour plots for this normalized coupling
  for $\mu=400GeV$ and $\tan\beta=10$. These curves are based on 
  the exact expression for the vertex (for example 
  see\cite{eenous_t1t1h}) including the one-loop corrections 
  to the mass and coupling of the Higgs. For clarity the $\mh$ and 
  $\Delta\rho$ constraint discussed above are reproduced there as well. In 
  Fig.~\ref{contth} one sees that $R_{\sto}$ cannot exceed 50. In fact 
the equipotential  $R_{\sto }=50$ almost coincides with the 
$\mh\ge 90GeV$ exclusion curve, thus it is the 
  maximum enhancement of coupling one can hope for. 
For certain values of the mixing angle, $\delta\rho$ excludes lower values of $R_{\sto}$. For example
   near the 
  maximal mixing, the $\Delta\rho$ constraint precludes $R_{\sto 
  }\ge 10$ while in the large $\cost$ region $R_{\sto 
  }$ could barely exceed 1.

When presenting our results we will, 
unless otherwise stated, impose the limits $\mh>90GeV, \Delta \rho 
<.0013$ \cite{Latest_mh_limit,Langacker_Fits98} together with the mild CCB
 constraint for $\mu=400GeV$, Eq.~\ref{CCBkusenko}.
 we also impose a limit on the squark mass, $\msbo\ge 80GeV$ 
 \cite{lep2squarklimit}.

\section{Results }

The calculation was performed with the use of the GRACE-SUSY 
package for automatic calculation of SUSY 
processes\cite{GraceSusy}. We modified the tree-level package to 
include the important radiative corrections to the Higgs mass  and 
couplings. We have included only one-loop corrections for the 
third generation squarks. For not too large values of $\tan\beta$,  
the stop contribution completely overwhelms the sbottom 
contribution. 
 As mentioned above, 
the relevant parameters are the masses of the stop squarks and the 
stop mixing angle.  The mass of the pseudoscalar is taken to be 
$M_A=1$~TeV while we have chosen $\mu=400GeV$. The latter 
parameter in principle enters  the $\sto \sto h$ vertex but in 
effect does not influence much the numerical results.
Although the $\sbt$  does not contribute to the $\sto\sto Z$ 
process, we  fixed $m_{\sbt}=800GeV$ to ensure that 
this particle cannot be directly produced even at $\sqrt{s}=800GeV$. 
 Due to the 
reduced  phase space available at a 500 GeV  collider, we have 
only considered the case $m_{\sto}=120$~GeV. For this mass,  the 
cross-section for $\epem\to\sto\sto Z$ can vary by  more than  an order of magnitude from 
$\approx .05fb - 1.5fb$ depending on the value of the input 
parameters as well as on the choice of polarisation.
Note that this is far from the orders of magnitude variations that
we encountered  for  $\sto\sto h$ production\cite{eenous_t1t1h}.
Fig.\ref{stostohz} shows how drastically $\sto\sto h$ changes as $\mstt$ is varied
compared to the mild variation of $\sto\sto Z$. The main reason for this difference
is that $\sto\sto h$ is completely dominated by the $\sto\sto h$ vertex whereas
in $\sto\sto Z$ different classes of diagrams contribute, Fig.\ref{feynman}.
 To get a 
better understanding on the dependence on the input parameters it 
is instructive to consider the contribution from each set of 
diagrams. The crucial point  to note is that some diagrams will 
involve only gauge couplings while others will involve Yukawa 
couplings. The latter are potentially large in the large mass splitting 
case. 

 \begin{figure*}[htbp]
\begin{center}
\mbox{\epsfxsize=14cm\epsfysize=12cm\epsffile{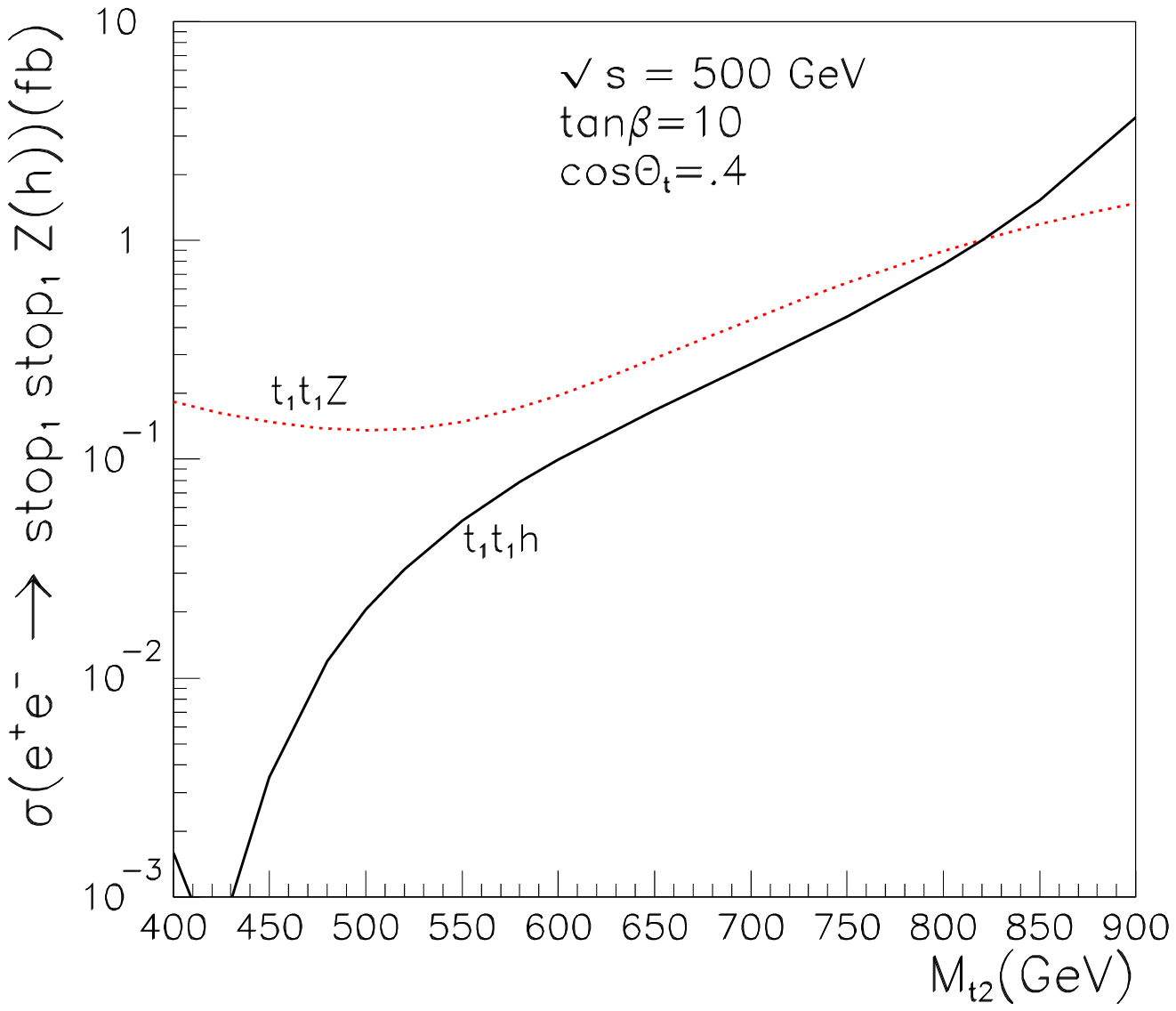}} 
\caption{\label{stostohz}{\em Comparison of $\epem \to \sto \sto 
h$ and $\epem \to \sto \sto Z$. $100\%$ right-handed $e^-$ polarisation
assumed and $\msto\approx 120GeV$. }} 
\end{center}
\end{figure*}

There are three classes of diagrams that enter this process,Fig.~\ref{feynman}, 

\begin{itemize}
\item{a)} Initial state Z radiation.
\item{b)}Final state Z radiation, this includes a diagram with a quartic vertex.
\item{c)}Final state Z radiation with exchange of a $\stt$.
\item{d)}  Higgs exchange diagrams. These also  include a diagram involving the
 exchange of the heavy Higgs, which however 
is  negligible. 
\end{itemize}

\begin{figure*}[htbp]
\begin{center}
\mbox{\epsfxsize=14cm\epsfysize=10cm\epsffile{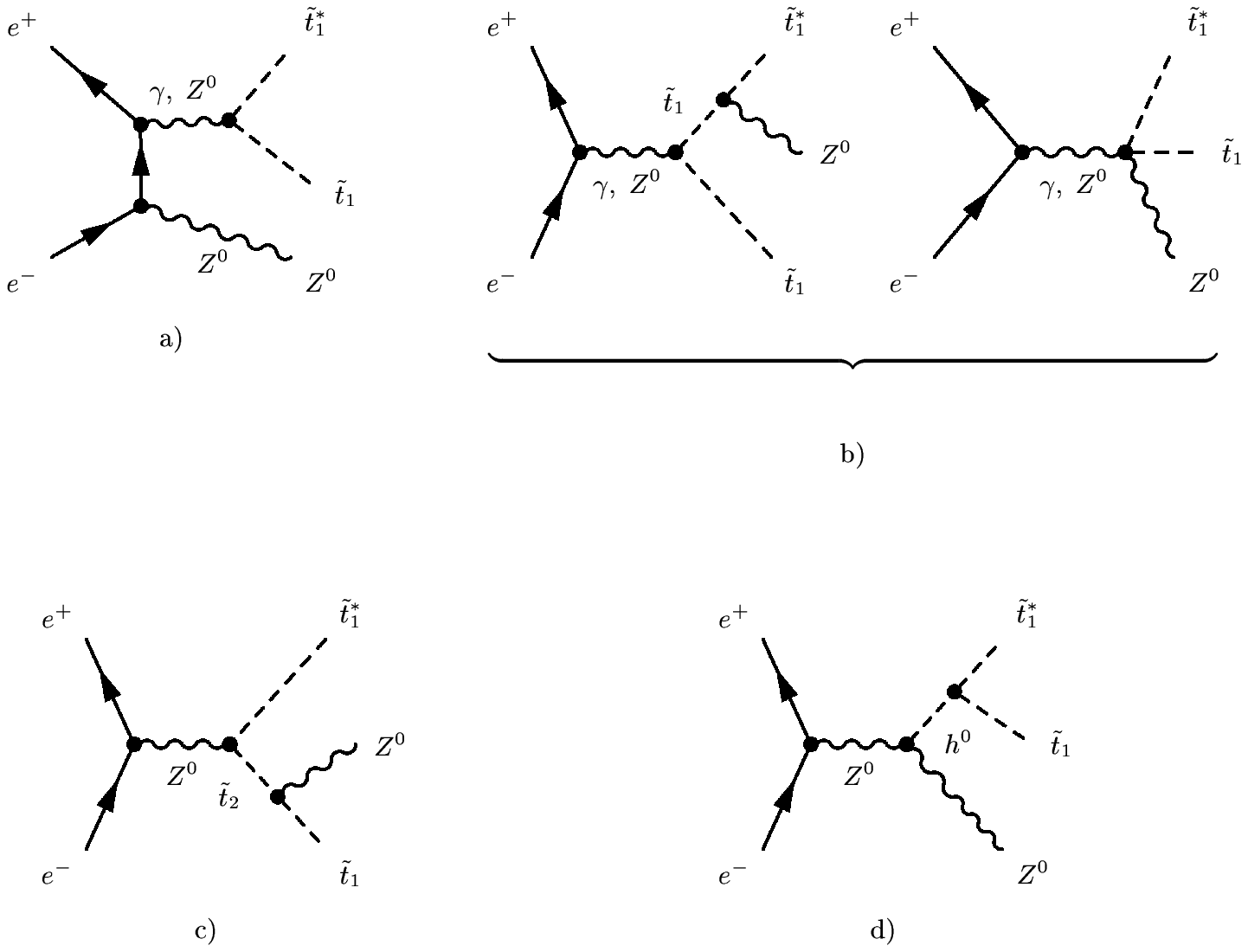}} 
\caption{\label{feynman}{\em Classes of Feynman diagrams for 
$\epem \to \sto\sto Z$. a) Initial state radiation b) Final state radiation
c) Final state radiation from $\stt$ d) Higgs exchange  }} 
\end{center}
\end{figure*}
 The only diagrams involving the potentially 
large Yukawa coupling  are, besides the Higgs exchange 
diagram, the ones corresponding to Z radiation off $\stt$ Fig.~\ref{feynman}c). 
The large Yukawa coupling arises from the Goldstone component of 
the coupling, thus when the splitting is large $\stt \ra \sto Z$ 
can be approximated by $\stt \ra \sto \phi^0$, $\phi^0$ being the 
neutral Goldstone Boson, with an effective coupling $g \frac{1}{4 
\mw} \sin 2\theta_{\stop} \; (\msttsq-\mstosq)=g/2M_W m_t 
(A_t+\mu/\tgb)$. Nonetheless these diagrams also have a 
$\frac{1}{m_{\stt}^2}$ factor from the propagator and we found the 
overall contribution to the cross-section to be rather small. Only 
the diagram with Higgs exchange will then feature a 
 Yukawa coupling enhancement, hence a $m_{\stt}$ dependence through the 
 $\sto\sto h$ coupling. 
 This diagram will contribute to the cross-section according to the
 strength of the $\sto\sto h$ vertex, from negligible to almost 100\%, as
 Fig.~\ref{higgsexchange} shows. In fact the contribution of this diagram can almost
 be inferred from the cross-section $\epem \to \sto\sto h$, see
 Fig.\ref{stostohz}.
  As for the Z radiation 
diagrams, they are dominated by the contribution from Z radiation 
off initial beams (an order of magnitude larger than the Z 
radiation off stops).  
They account for $\sigma=.2fb$ at $\cost=0.4$. For a 
 collider of luminosity ${\cal L}=500 fb^{-1}$,  this corresponds to 
 over $100$ raw events. While these events could be recorded and 
 the cross-section measured, it would not provide any additional 
 information on the value of the unknown parameter of the stop 
 sector, this could be considered as  "background" events.

\begin{figure*}[htbp]
\begin{center}
\mbox{\epsfxsize=14cm\epsfysize=11cm\epsffile{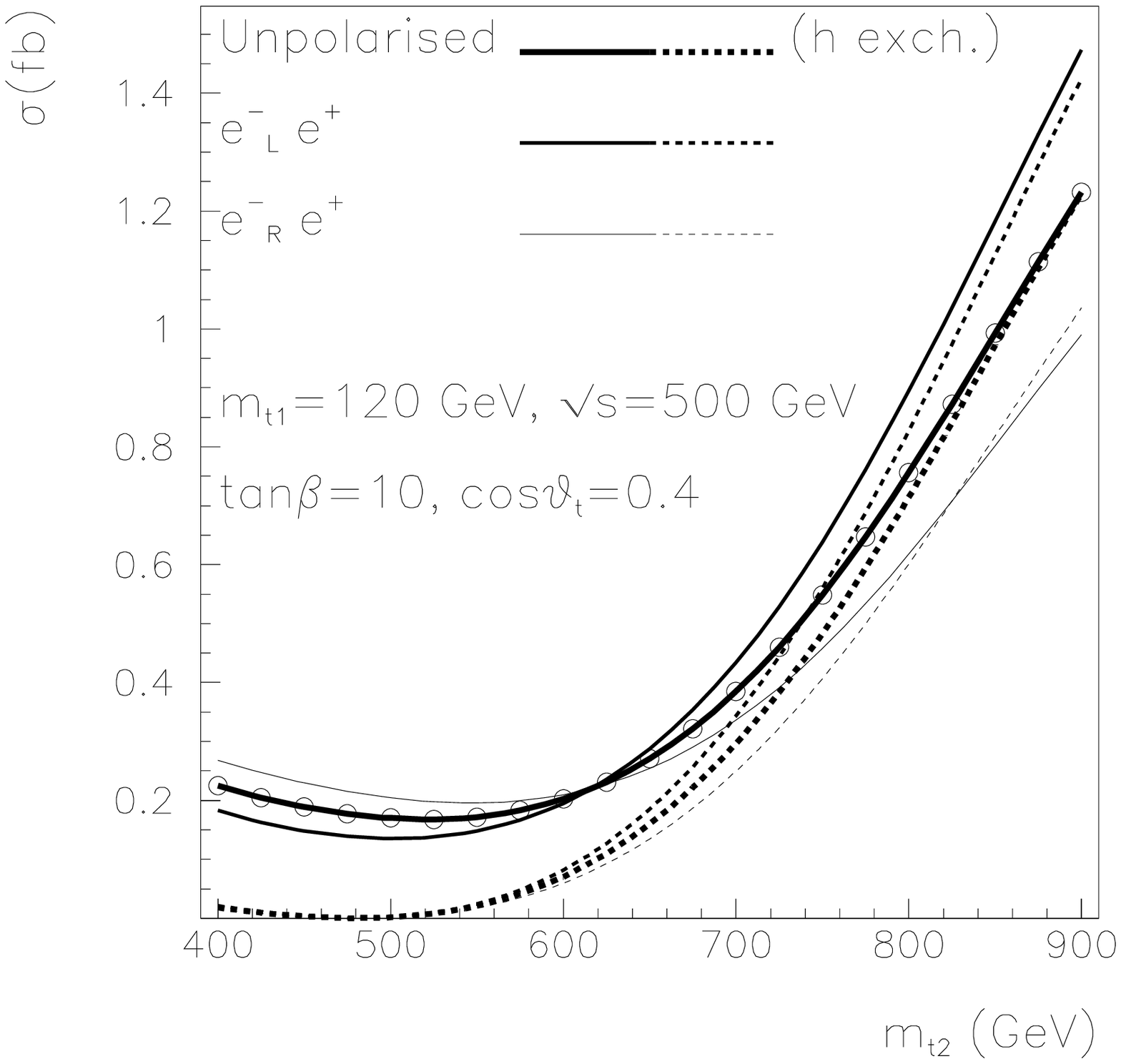}} 
\caption{\label{higgsexchange}{\em Polarized and unpolarized 
cross-section for $\epem \to \sto\sto Z$, $\cost=0.4$, $\tgb=10$ 
and $\mu=400GeV$. The contribution of the 
Higgs exchange diagram is displayed(dash).  }} 
\end{center}
\end{figure*}

To analyse the $m_{\stt}$ dependence of the cross-section, 
first 
consider the case of intermediate mixing,  for example $\cost=.4$.  
As was the case for $\epem \to \sto\sto h$, the cross-section is  
smallest for $m_{\stt} \approx 400-600GeV$, this corresponds to  
the region where the 
 $\sto\sto h$ vertex drops significantly. 
 As the $\stop_2$ mass increases the cross-section increases significantly
 by almost one order of magnitude.
  As 
reflected in Fig.\ref{higgsexchange}, this is essentially due to 
the rapidly rising contribution from the Higgs exchange diagram, 
itself driven by the coupling $\sto\sto h$. At $m_{\stt}=900$GeV, 
the Higgs diagram alone explains the major part of the 
cross-section, although some important interference effect between 
the Higgs exchange diagram and all other  diagrams remains. In 
particular  there is some small constructive interference between 
the initial Z breamstrahlung and the Higgs contribution and a more 
important (at the 10\% level ) destructive interference between 
the final breamstrahlung and the Higgs exchange diagram. For this 
particular value of the mixing angle, the  various contributions  
conspire to cancel each other at the highest mass and one is left 
with a cross-section which seems nearly 100\% arising from the $h$ 
exchange diagram. This fortuitous cancellation at $\cos\theta_{\stop}=0.4$
does not occur when 
one looks at the polarised cross-section or at any other value of 
the mixing angle. Note that this  behaviour   is in stark contrast  
with what was obtained for $\epem \to \sto \sto h$. There, 
whenever the $\sto\sto h$ vertex vanishes, the cross-section becomes exceedingly 
small, since the only diagram that does not contain either $\stop\stop 
h$  vertices, the one originating from $\epem \to hZ $ 
is completely negligible for the whole range of parameters.
Indeed, at high energy a 
longitudinal Z, which is essentially a Goldstone boson, 
would be mainly produced but
 this Goldstone boson does not couple to $\sto\sto$.
On the other hand, in associated Z production , the same $hZ$ initiated diagram
gives a significant contribution to the cross-section as it is now the Higgs
accompanying the longitudinal Z
that splits into $\sto\sto$ pairs, and this with a potentially
large vertex enhancement.

Next  consider the effect of polarisation. 
 While for  
$\sto\sto$ pair production the value of the stop mixing angle determined 
the  polarised cross-section, for the  $\sto \sto Z$ process one 
has to take into account other parameters as well. In the  region where
 the cross-section arises mainly from the 
diagrams with a Z  breamstrahlung, either initial or final,  the 
polarisation dependence is  expected to be similar to the one 
for stop pair production since essentially gauge couplings, which do not change the chirality are involved. In the latter process,  the 
cross-section is dominated by $e^-_R$ for $\cost \le .5$ otherwise 
by $e^-_L$. However when the Higgs coupling to $\sto\sto$ becomes 
large, which for intermediate or large mixing corresponds to the
 large $m_{\stt}$ region,
   it is the Higgs exchange diagram that is responsible for 
most of the cross-section, Fig.\ref{higgsexchange}. In this case the 
dominant polarisation configuration is the same as for  an 
s-channel Z production. The ratio of the polarised cross-section 
is given more or less by the ratio of the couplings of the Z to 
$e_L$ and $e_R$ respectively.
Note that the 
difference between the two polarised cross-sections is not very pronounced for 
the value of the mixing we have chosen, it is  more marked 
in the case of small mixing, $\sintt\approx 0$. 
Furthermore for large $\cost$ one expects $e^-_L$ to dominate whether or not
one benefits from the large Yukawa enhancement.  The expectations for
different values of $\cost$ will be discussed next.

For  $\cost\approx 0$, $\sin 2\theta_{\stop}\approx 0$, one expects 
 from the expression of the $\sto\sto h$ 
vertex, Eq.~\ref{approxtth}, a very mild dependence on 
the $\stt$ mass, see Fig.\ref{costheta}.
As a result of the $\mt^2$ contribution cancelling against the mixing 
contribution, as $\cost$ increases the strength of the $\sto\sto h$ vertex 
decreases until, for $\cost=.2$ and $m_{\stt}=900GeV$, there is a 
precise cancellation between the stop/top term in the vertex.
 One is then  left with only the contribution 
from the breamstrahlung diagrams.  
 Only when the mixing becomes 
significant can one see a rise in the cross-section at large 
masses. As discussed in the previous section,
 the $\Delta\rho$ constraint eliminates the upper 
range of cross-section as indicated by dots in Fig.\ref{costheta}.
In fact when mixing reaches $\cost=0.6$,  the maximum value for 
$\mstt\approx 600GeV$ and $\sigma\le 0.4 fb${\footnote{
In the maximal mixing region one hits a non-physical region
where the Higgs mass is driven 
negative and the cross-section cannot even be computed, as is the 
case  for example for $0.42\le\cost\le 0.88$ when $\mstt=900GeV$.}}. 
 For large $\cost$, because of 
the $\Delta\rho$ constraint (see Fig.\ref{contth}) one does not benefit from  the strong enhancement 
of the vertex and $\sigma$ cannot exceed $0.3fb$. Furthermore, 
in these configurations the sbottom is often directly accessible 
in the pair production process, this is indicated by an arrow in 
Fig.\ref{costheta}. The numerical results confirm what we had 
anticipated in the previous section, whenever the sbottom can be 
pair-produced, there is not much interest in the 3-body processes. 
This point concerns not only the  $\sto\sto Z$ but also $\sto\sto h 
$ production as this is just a reflection of the strength of the 
$\sto\sto h$ vertex. 

The polarised cross-sections follow approximately the same pattern,
see Fig.~\ref{costhetapol}.
As expected, the $e^-_L$ is dominant for large $\cost$,
the cross-section can reach 0.75fb even for a low mass
$m_{\stt}=400 GeV$ at $\cost=.9$. The same polarisation dominates also
 for intermediate
$\cost$ provided $m_{\stt}$ is large, that is large Yukawa coupling,
otherwise the choice of a  right handed electron polarisation gives a
larger cross-section.

For all numerical results presented we have taken $m_{\stt}\le 900GeV$, the
maximum value allowed for $\cost=0.4$. However one should keep in mind that for
smaller values of $\cost$, the ``large" Yukawa enhancement of the 
 cross-section occurs for $\stt$  masses above 1TeV
 which still passes all constraints,
  see Fig.~\ref{constraint}.  
Nevertheless for these angles the fluctuations with $m_{\stt}$ are
never dramatic and lie within  the $3\sigma$ interval 
 with a high-luminosity ${\cal L}=500 fb^{-1}$.

\begin{figure*}[htbp]
\begin{center}
\mbox{\epsfxsize=14cm\epsfysize=12cm\epsffile{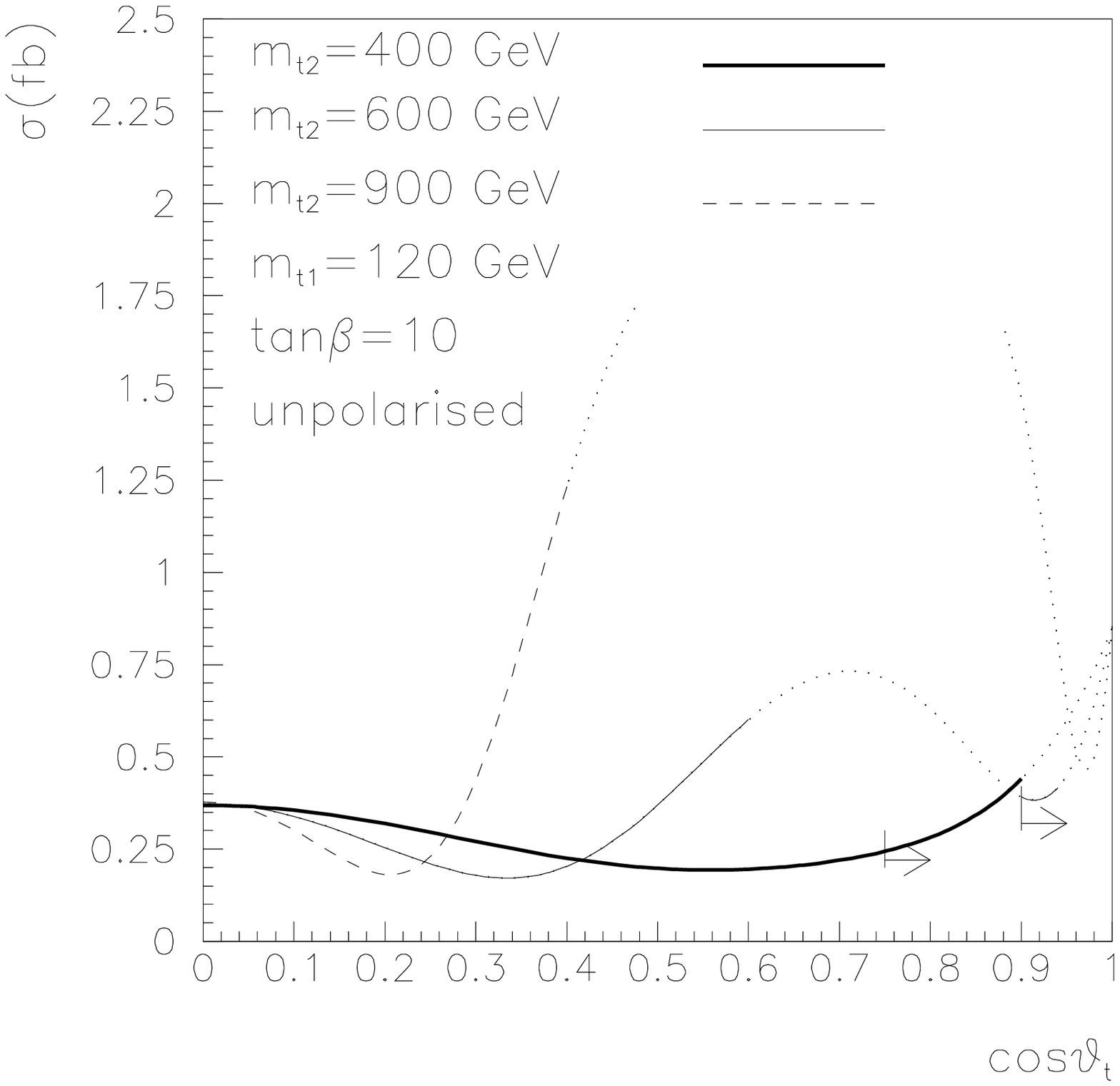}} 
\caption{\label{costheta}{\em $\sigma(\epem\sto\sto Z)$ vs $\cost$ 
for $m_{\sto}=120 GeV$ $m_{\stt}=400,600,900 GeV$. Points that do 
not pass the constraints are indicated as dots. To  the right of the arrows, 
$\sbo$ pair production opens up ($m_{\sbo}\le 250 GeV)$. }} 
\end{center}
\end{figure*}
 
\begin{figure*}[htbp]
\begin{center}
\mbox{\mbox{\epsfxsize=8cm\epsfysize=10cm\epsffile{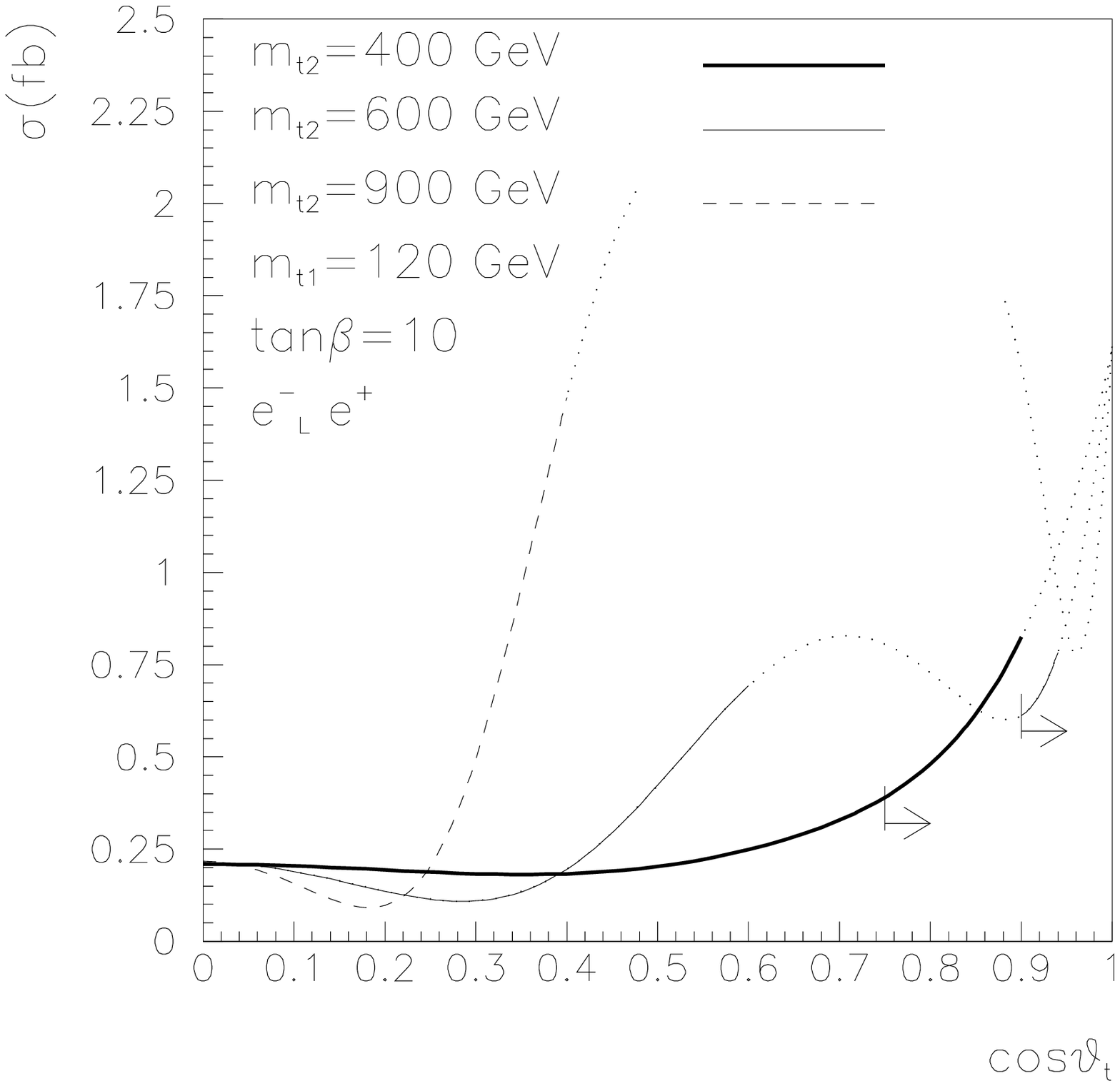}} 
\mbox{\epsfxsize=8cm\epsfysize=10cm\epsffile{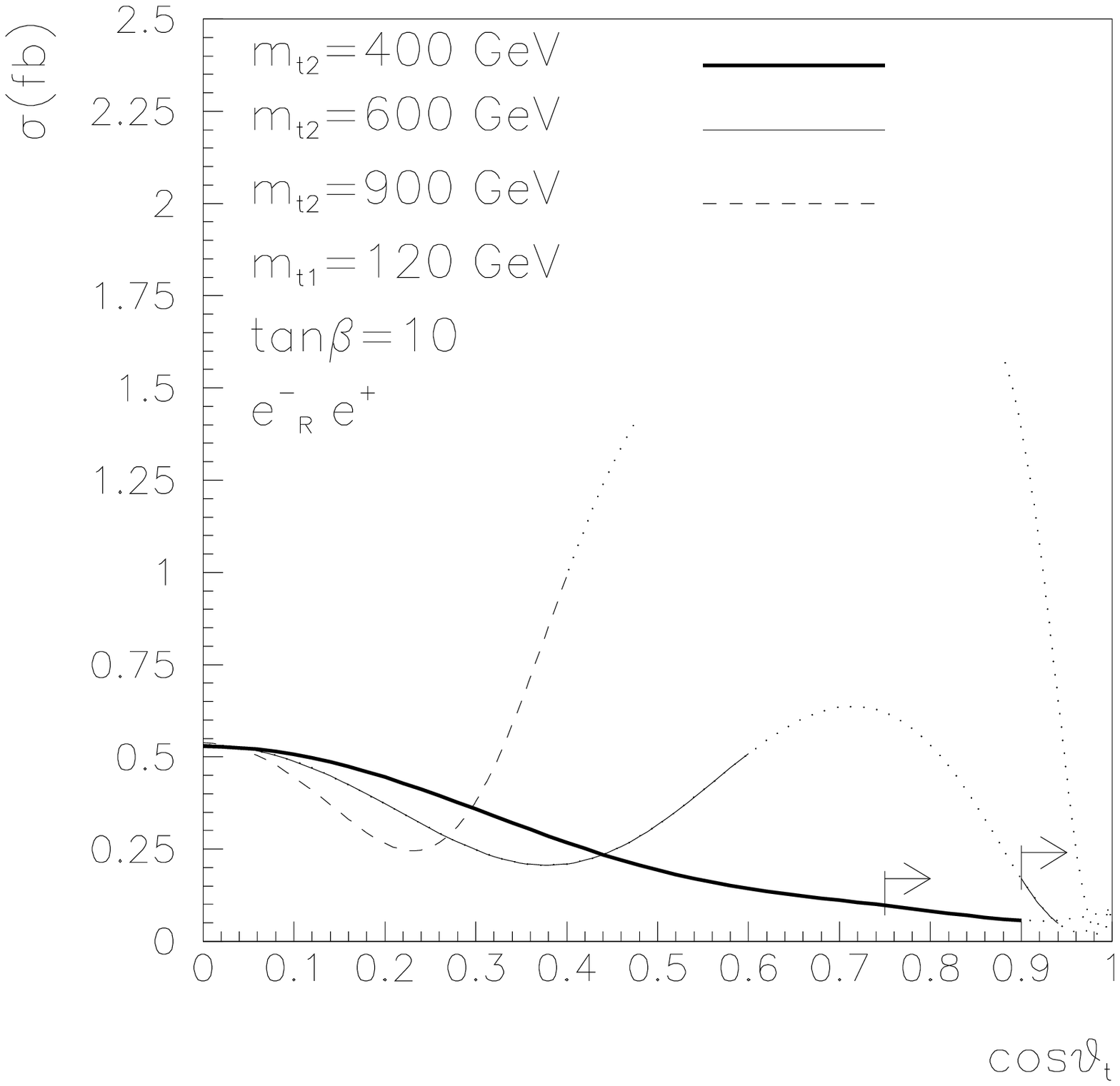}} 
} 
\caption{\label{costhetapol}{\em $\sigma(\epem\to \sto\sto Z)$ vs 
$\cost$ for $m_{\sto}=120GeV$, $m_{\stt}=400,600,900$GeV with polarised beams. The meaning of the dotted lines and the arrows is the same as in the previous figure.}} 
\end{center}
\end{figure*}

We have already alluded to some of the  differences between associated Higgs and
associated Z production. The main point is that  one does not expect as sharp a  
 dependence on the  $\sto \sto h$ vertex (that is on $m_{\stt}$ for a given
 mixing angle)  as the process $\sto \sto h$ itself
 since only one diagram 
contributing to the $\sto \sto Z$ cross-section involves the 
Higgs. However as 
 compared to the latter the associated Z channel features a larger
 cross-section for a large range of parameters. It could therefore
 be used in conjonction with the associated Higgs channel to help
 determine the parameters of the stop sector, in particular the 
 mass of the $\stt$, Fig.~\ref{stostohz}. For example, assuming an 
 efficiency of $50\%$ and an intermediate value for 
 $\sigma=.38fb$ at $\cost=.4$ for unpolarised beams
 one could deduce from a $3\sigma$ measurement a mass 
 $\mstt=700^{+30}_{-50}GeV$. The uncertainty is of the same order as that 
 expected in $\epem\to\sto\sto h$\cite{eenous_t1t1h}. For this particular mixing angle,
 roughly the same precision is expected from either electron beam
 polarisation.

 If the lightest
stop turns out not to be so light, one would need to go to higher centre-of-mass  
energies to observe some events from  the associated production of 
stop and Z. However higher energies can also mean more phase space 
for the direct production process $\epem \to \sto\stt$.
 Not only 
could the  direct production of $\stt$ allow for the determination 
of  $m_{\stt}$ it  can also trigger the final state $\sto\sto Z$. 
This occurs when the $\stt$  further decays in $\sto Z$. 
Cross-sections of a few  $fbs$ can  be reached\cite{viennast2st1}, and 
the partial width into this mode can 
be quite large, since the $\stt \to \sto Z_L$ can benefit from the 
large Yukawa enhancement, as discussed earlier. This branching fraction however 
 depends on   the parameters of the MSSM and in 
particular those of the sbottom sector. We will not entertain 
this possibility any longer,   here we rather consider only values 
of the two physical stop masses such that $\sto\stt$ is above threshold 
. We have considered $\sqrt{s}=800GeV$ and different values of $\msto$ while 
varying $\mstt$ in the range such that $\sto\stt$ is above threshold.
 In this case we only see a mild
dependence on the  $\sto\sto h$ vertex, and that mostly for the 
lower values of $m_{\sto}$, Fig.~\ref{ttz800}. 
At this energy one can hope for a signal only for $m_{\sto}$ below about 250GeV.

\begin{figure*}[htb]
\begin{center}
\mbox{\epsfxsize=14cm\epsfysize=12cm\epsffile{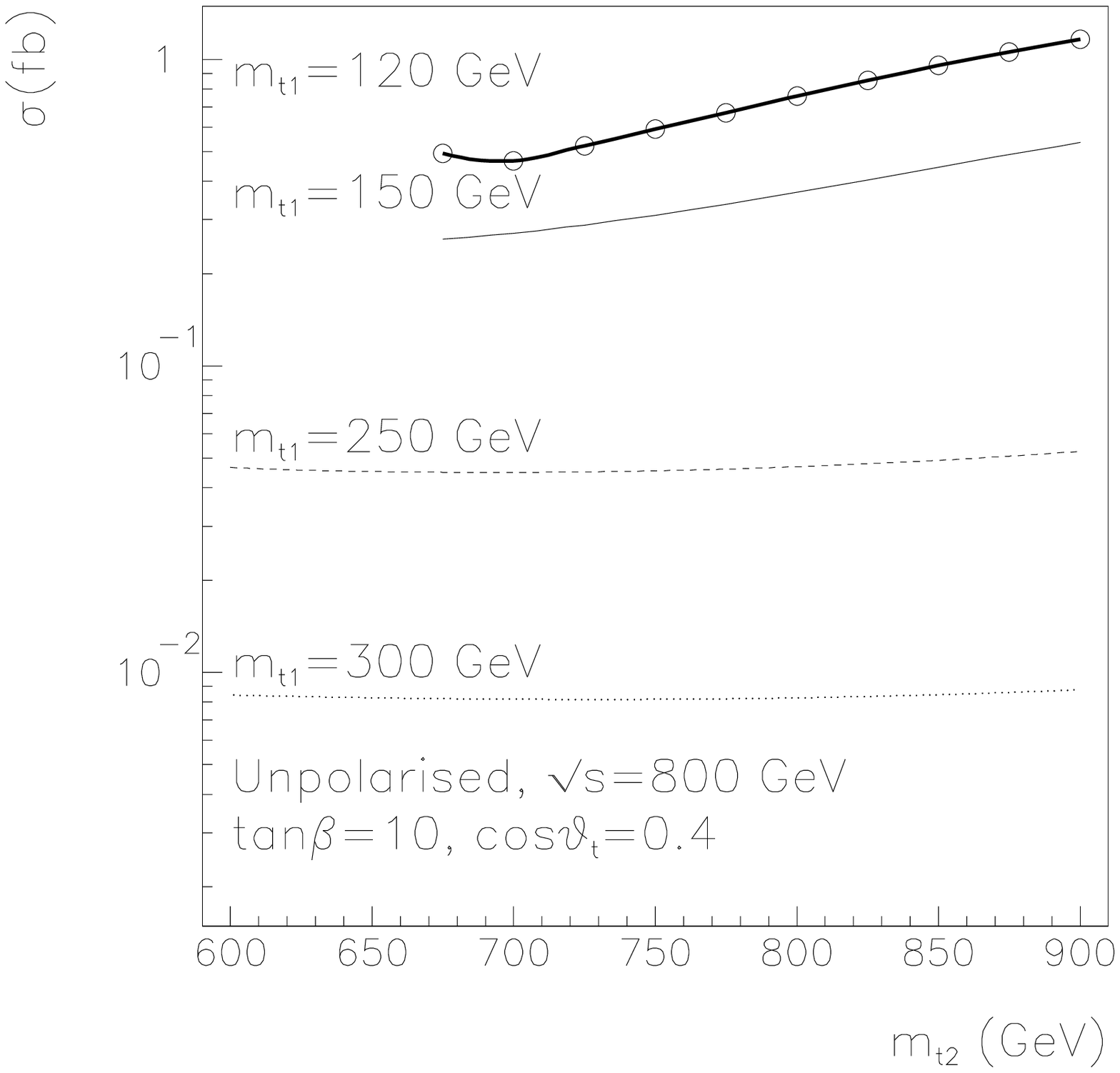}} 
\caption{\label{ttz800}{\em 
Cross-section for $\epem \to \sto\sto Z$ at 800GeV for
$m_{\sto}=120,150,250,300GeV$,  $\cost=0.4$, $\tgb=10$ 
and $\mu=400GeV$. }} 
\end{center}
\end{figure*}

An important issue that remains to be quantified is the detectability of the
signal both for the associated Higgs and the associated Z processes. For the
parameters we are considering here, where beside $\sto$ only h and the LSP are light, the only decay mode of $\sto$ is into
$c\chi^0$. An analysis of signatures and background for the stop pair production 
including the decay mode we are considering already exists\cite{bartl}. This 
issue is  also important for the extraction of mass and mixing angle
in the pair production. Furthermore for the precise measurements of 
these parameters the question of  radiative corrections
  needs to be taken into account.
  All the results presented here correspond to a rather large value for $\mu$
  ($\mu=400GeV$), 
different values of $\mu$
could lead to a different MSSM spectrum and eventually different decay modes for the $\sto$.
This could even facilitate the extraction of the signal. However  
for the production process itself,  the numerical results
would not differ much in the region that is most interesting, the large 
mass splitting region,
since the contribution from the $\mu$ term to the vertex is small compared with the trilinear coupling
contribution. Here we have used only one-loop corections to  $\mh$, the inclusion 
of the dominant two-loop corrections\cite{higgsmass_twoloop_exact} 
should not affect the results  very much as in associated Z production, 
we have found very little dependence on the precise value of the Higgs mass.

In conclusion, the process $\epem \to \sto\sto Z$ should be measurable at 
a high luminosity linear
collider such as TESLA, provided the $\sto$ is not very far above the present limit. 
While observable for any values of the parameters,  
(our conclusion applies to the small $\tan\beta$ regime), this
process can give additional information on the not directly observable $m_{\stt}$
provided there is large mixing and mass splitting. In this sense it is very similar to
$\sto\sto h$.  There are regions in parameter space where only the $\sto\sto Z$ would be
observable. In this worst-case situation, even though  the cross-section does not depend 
strongly on the
parameters,  one can still get a rough determination of a range for $m_{\stt}$ . For
the intermediate mixing we have discussed at length, non observation of $\sto\sto h$ and
a ``low" value for  $\sto\sto Z$
would indicate a $\stt$ in the 400-600 GeV range.

\end{document}